\documentclass[
letter,
twocolumn,
superscriptaddress,
amsmath,
amssymb,
prl,
nofootinbib
showkeys,
10pt,
floatfix
]{revtex4-2}

\usepackage{graphicx}
\usepackage{dcolumn}
\usepackage{bm}
\usepackage{xcolor}
\usepackage{ulem}
\usepackage{hyperref}
\usepackage{cleveref}
\usepackage{amsmath}
\usepackage{float}

\usepackage{svg}
\usepackage{amssymb}
\usepackage{scalerel}
\usepackage{caption}
\usepackage{setspace}
 \usepackage{multirow}
 \usepackage{booktabs}
 \usepackage{siunitx}
 \usepackage{physics}
 \usepackage{amsmath}

\makeatletter
\pretocmd\frontmatter@keys@format{\addvspace{20\p@}}{}{}
\makeatother

\usepackage[font=small]{caption}

\begin{document}

\title{
Purifying single photon emission from giant shell CdSe/CdS quantum dots at room temperature
}

\author{Sergii Morozov}
\affiliation{
 Center for Nano Optics, University of Southern Denmark, Campusvej 55, DK-5230 Odense M, Denmark
}

\author{Stefano Vezzoli}
\affiliation{
 The Blackett Laboratory, Department of Physics, Imperial College London, London SW7~2BW, United Kingdom
}

\author{Alina Myslovska}
\affiliation{
Department of Chemistry, Ghent University, Krijgslaan 281-S3, Gent 9000, Belgium
}

\author{Alessio Di~Giacomo}
\affiliation{
Department of Chemistry, Ghent University, Krijgslaan 281-S3, Gent 9000, Belgium
}

\author{N.~Asger~Mortensen}
\affiliation{
 Center for Nano Optics, University of Southern Denmark, Campusvej 55, DK-5230 Odense M, Denmark
}
\affiliation{
 Danish Institute for Advanced Study, University of Southern Denmark, Campusvej 55, DK-5230 Odense M, Denmark
}

\author{Iwan Moreels}
\affiliation{
Department of Chemistry, Ghent University, Krijgslaan 281-S3, Gent 9000, Belgium
}%

\author{Riccardo Sapienza}
\affiliation{
 The Blackett Laboratory, Department of Physics, Imperial College London, London SW7~2BW, United Kingdom
}%
\email{r.sapienza@imperial.ac.uk}

\begin{abstract}

Giant shell CdSe/CdS quantum dots are bright and flexible emitters, with near-unity quantum yield and suppressed blinking, but their single photon purity is reduced by efficient multiexcitonic emission. We report the observation, at the single dot level, of a large blueshift of the photoluminescence biexciton spectrum ($24\pm5$~nm over a sample of 32 dots) for pure-phase wurtzite
quantum dots. 
By spectral filtering, we demonstrate a 2.3 times reduction of the biexciton quantum yield relative to the exciton emission, while preserving as much as $60\%$ of the exciton single photon emission, thus improving the purity from $g_2(0)=0.07\pm0.01$ to $g_2(0)=0.03\pm0.01$. 
At larger
pump fluency the spectral purification is even more effective with up to a 6.6 times reduction in
$g_2(0)$, which is due to the suppression of higher order excitons and shell states experiencing even larger blueshift.
Our results indicate the potential for synthesis engineered giant shell quantum dots, with further increased biexciton blueshift, for quantum optical applications requiring both
high purity and brightness.

\end{abstract}


\maketitle

\section{Introduction}

\begin{figure*}
	\includegraphics[width=\linewidth]{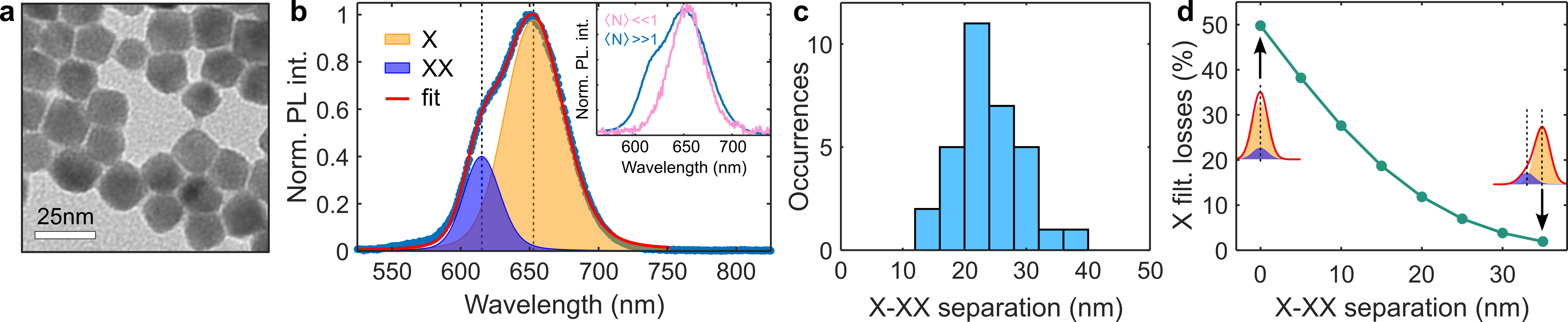}
	\caption{\textbf{Giant shell CdSe/CdS colloidal quantum dots with a large biexciton blueshift.}	
	\textbf{a}~TEM image of pure-phase wurtzite quantum dots with a 3.4~nm CdSe core and a thick CdS shell consisting of ca.~20 CdS monolayers. 
	\textbf{b}~Photoluminescence spectrum of an individual quantum dot above saturation $\langle N \rangle \gg 1$. The fit (red curve) is the sum of two Gaussian peaks, one for the single exciton (labeled X, orange) and one for the biexciton (XX, blue) contribution.  The inset compares spectra measured below (pink) and above (blue) saturation, highlighting the appearance of a blueshifted peak at high pump fluency.
	\textbf{c}~Statistics of the biexciton blueshift in 32 individual quantum dots, measured as the difference in nm between the X and XX peaks.
	   {\textbf{d}~Simulations of single exciton filtering losses at increasing biexciton blueshift (see text for details).}
	}
	\label{fig1}
\end{figure*}

The development of blinking-free, bright and pure single photon emission from quantum dots is very important for quantum cryptography and computing~\cite{Kagan2020}. Colloidal quantum dots hold potential as room-temperature single-photon emitters with widely tunable optical properties. A quantum dot is an isolated quantum system, whose {  properties --} such as exciton emission wavelength, lifetime and quantum yield {  --} can be controlled by engineering the quantum dot shape and morphology, via the electron and hole wavefunctions design~\cite{Murray2000,Bailey2003}. 

In contrast to cryogenic quantum dots -- where the neutral excitonic emission can be easily spectrally filtered~\cite{Zwiller2005} -- colloidal quantum dots, at room temperature, have a broad excitonic emission spectrum which overlaps with multiexcitonic bands. Efficient Auger recombination ensures single photon emission in colloidal quantum dots, by suppressing many-body emission, however, it also leads to the excitation of charged and {   trapped} states, often associated with photoluminescence blinking and reduced brightness~\cite{Efros2016}. 

Giant shell nanocrystals can significantly reduce Auger recombination and protect excitons from surface interactions, therefore reducing blinking and increasing quantum yield~\cite{chen2008,Bae2013}. However, this implies that stable, neutral multiexcitons can efficiently recombine, leading to a reduced emission purity, and a typical intensity auto-correlation function at zero delay of $g_2(0)>0.3$~\cite{Fisher2005PRL,Park2011}. 

The simplest strategy to reduce multiexciton contributions in colloidal dots is to set the pump fluency well below saturation~\cite{Klimov2007}. This, however, limits drastically the quantum {  dots'} brightness~\cite{Nasilowski2015} and their ability to produce single photons on demand. Alternatively, temporal, polarization and spectral filtering can increase the purity. In temporal filtering, photons with a fast decay time can be removed after time-tagging their arrival time, with $g_2(0)$ as low as 0.005 at room temperature~\cite{Ihara2019}. However, the temporal filtering is based on post-processing, which is not suitable for all applications~\cite{Mangum2013}. 
Polarization and spectral filtering of unwanted multiexciton contributions is so far restricted to {cryogenically operated} quantum emitters, because of broadened emission lines and polarization decoherence at room temperature~\cite{Wang2019}. In order to make spectral filtering effective at room temperature, biexciton separation of the order of the exciton bandwidth is required.

In recent years giant shell CdSe/CdS dot-in-rod and rod-in-rod nanocrystals, which are pure-phase wurtzite core/shell, have shown evidence of a large biexciton blueshift~\cite{Christodoulou2015,Polovitsyn2018}, {   exceeding blueshifts in } commonly studied {  zincblende}/wurtzite quantum dots~\cite{2Galland2012}. Calculations of the {  electronic} band structure have revealed that in pure-phase wurtzite nanocrystals with a thick CdS shell, the strain at the CdSe/CdS interface yields a strong piezo-electric field across the CdSe nanocrystal core~\cite{Christodoulou2015,Segarra2016}, and in turn an enhanced electron-hole wavefunction separation and a large blueshift of the biexciton emission. 

Here we study pure-phase wurtzite CdSe/CdS giant shell quantum dots, synthesized with oleic acid so that the shell grows symmetrically in all directions, and we characterize their optical properties at the single dot level. We report an average exciton quantum yield of $50\pm5\%$ and a lifetime of $638\pm356$~ns over a sample of 32 dots. 
More importantly, we observe a large biexciton blueshift {  of} $24\pm5$~nm, which is of the same order of the exciton spectral width ($40\pm5$~nm). 
We exploit this large shift to spectrally filter the single photon emission and show a 2.3 times reduction of the biexciton quantum yield relative to the exciton emission, which improves single photon purity from $g_2(0)=0.07\pm0.01$ to $g_2(0)=0.03\pm0.01$. 
We also show that spectral filtering becomes more effective at larger pump fluency, just above the exciton saturation, with up to a 6.6 times purity improvement. An average sample of 16 quantum dots {  --} pumped just above saturation and filtered above 650~nm {  --} is shown to undergo a four-fold reduction in $g_2(0)$ (0.34 to 0.08) with only a 1.85 reduction in brightness (50$\pm5\%$ to 27$\pm3\%$). {  This is} showing the potential of spectral filtering to strongly reduce the contribution of multiexcitons as well as parasitic shell states, while preserving most of the exciton emission.

\section{Results and Discussion}

\subsection{Giant shell quantum dots}

We synthesize colloidal CdSe/CdS giant shell quantum dots following a protocol reported in~\cite{Christodoulou2014} with slight modifications (see SI for more details): we use oleic acid instead of phosphonic acid, so the growth of the CdS shell is symmetric rather than happening mainly in one direction, which would create rod-like shells.
The quantum dots have a 3.4~nm CdSe core covered by a thick shell consisting of about 20 CdS monolayers, resulting in a total diameter of $17\pm3$~nm (see Fig.~\ref{fig1}a for a TEM image).
We use a pulsed blue laser ($1/T_{\rm rep}=$1~MHz, 404~nm) to locate and excite individual quantum dots in a scanning confocal microscope. 
Due to the giant shell, absorption occurs preferably in the CdS shell with a consequent nonradiative relaxation of carriers to the CdSe core, where radiative recombination occurs. The excitation power is converted to the number of generated electron-hole pairs $\langle N \rangle$, or excitons, inside the dots, according to a well established procedure, which is detailed in the following section. 

The radiative relaxation of generated excitonic and multiexcitonic states at room temperature results in a photoluminescence emission centered around 650~nm with an average full-width-at-half-maximum (FWHM) of 40~nm (see SI {Fig.~S}1 for the emission statistics over 32 quantum dots    {and {Fig.~S}2 for photo-stability analysis}). The measured neutral exciton emission lifetime extends to hundreds of ns ($638\pm356$~ns, see SI for more details), in line with previous observations on pure-phase wurtzite giant shell quantum dots~\cite{Christodoulou2014,Morozov2020}, and surpassing by almost an order of magnitude that of previously reported {  zincblende}/wurtzite giant shell quantum dots~\cite{Park2011,Bae2013}. This difference is mainly due to the additional buildup of a piezo-electric field across the CdSe core, which localizes the electron and hole wave-functions at opposite {  surfaces} of the quantum dot core and thus leads to a significant reduction of the electron-hole overlap~\cite{Segarra2016}.

The photoluminescence spectrum shows a blueshifted emission at high excitation power, see the inset in Fig.~\ref{fig1}b for a representative quantum dot, which is mainly due to the contribution of efficient biexciton emission~\cite{Park2011,Bae2013}. Single exciton emission dominates the spectrum of an individual quantum dot at very low pump fluency ($\langle N \rangle \ll 1$, pink line), while the spectrum at high pump fluency has a blueshifted shoulder ($\langle N \rangle \gg 1$, blue line). 
Fig.~\ref{fig1}b shows an agreement with a two-Gaussian fit at high fluency, the orange curve being the exciton (X) contribution and the blue one the biexciton (XX); the relative weight of the two spectra (i.e. their total area) is fixed by auto-correlation and lifetime measurements in order to improve the accuracy of the fit (see SI {Fig.~S}3 for details). We attribute the biexciton spectral blueshift to the enhancement of repulsive interactions among multiple electrons in the conduction band and among holes in the valence band, brought about by the localization of electrons and holes at opposite {  surfaces} of the core due the strong piezo-electric field~\cite{Christodoulou2015,Polovitsyn2018}.  
The {  spectral} shoulder further to the blue, around 570~nm, is evidence of higher order multiexciton emission and parasitic shell emission~\cite{Fisher2005PRL,2Galland2012}.
We characterize the magnitude of the biexciton blueshift by extracting the difference between the maxima of the blue{  -filled} and orange{  -filled} peaks obtained by fitting the spectra of 32 individual quantum dots. The resulting statistics is summarized in Fig.~\ref{fig1}c and shows a distribution with an average value and standard deviation of $24\pm5$~nm ($78\pm15$~meV). 
{Fig.~\ref{fig1}d presents simulations of expected losses for single exciton introduced by the spectral filtering. For the sake of simplicity, we perform a simulation of how much of the single exciton emission one needs to sacrifice to filter 50\% of the biexciton emission as a function of their spectral separation in a given batch of quantum dots. Here we used typical emission parameters of quantum dots: Gaussian spectral profiles for X emission maximum at 660~nm with FWHM of X and XX emission of 40~nm. The obtained results in Fig.~\ref{fig1}d demonstrate the minimal ($<$ 12\%) single exciton losses already for only 20~nm X--XX separation.}

\begin{figure}
		\includegraphics[width=0.99\columnwidth]{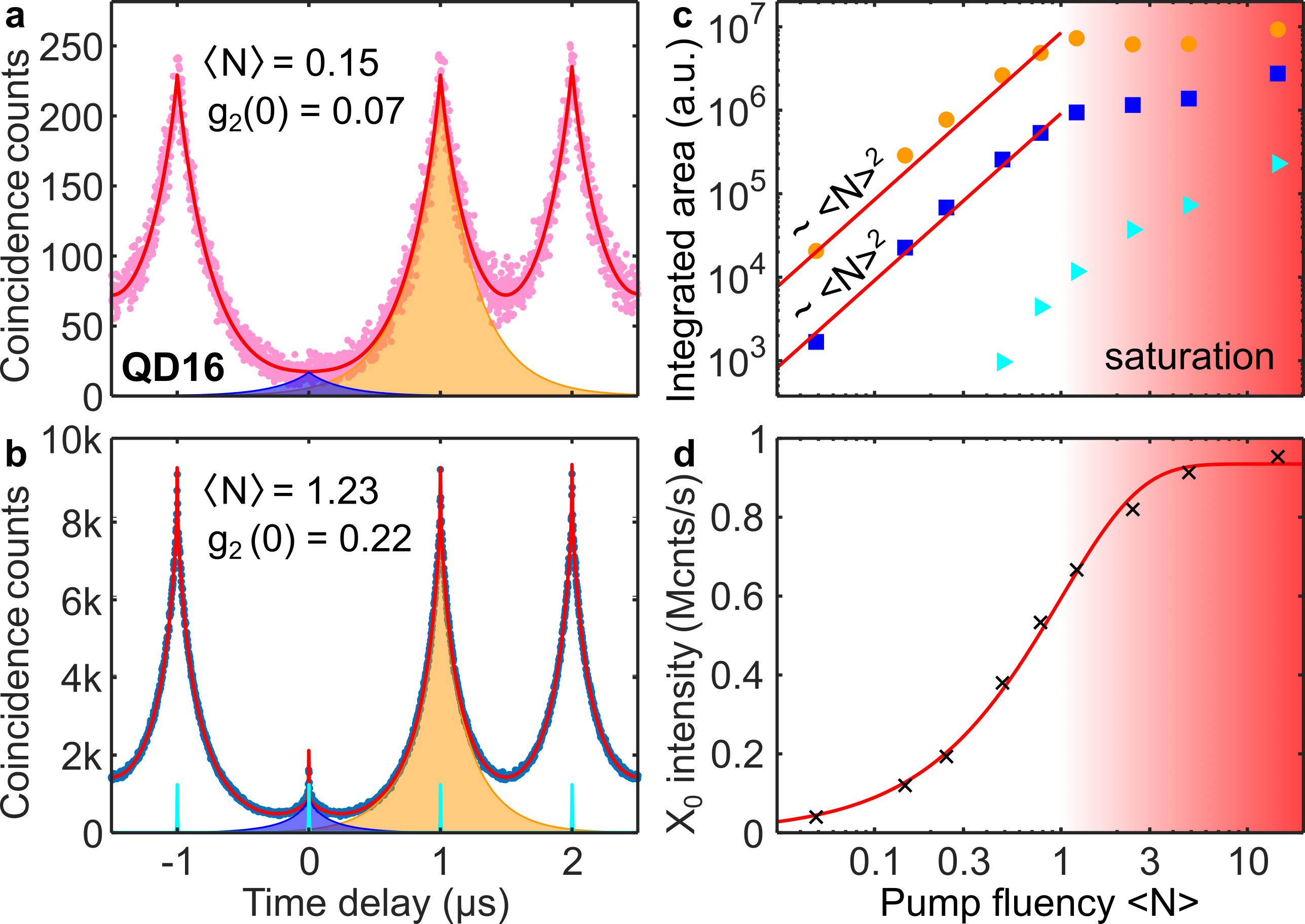}
	\caption{\textbf{Response of an individual quantum dot to increasing pump fluency.}
	\textbf{a}~Intensity correlation histogram (pink) measured well below intensity saturation. The central (blue) and side (orange) peaks are extracted from the fit (red) of the histogram.
	\textbf{b}~Intensity correlation histogram measured at high pump fluency, revealing a short-lifetime parasitic contribution of the CdSe shell emission (cyan peaks).
	\textbf{c}~Integrated areas of the central (blue squares) and side (orange circles) peaks at increasing pump fluency. Cyan triangles represent the emission of shell states. The saturation regime is indicated by the red-shaded area. At low fluency both the central and side peak areas grow as $\langle N \rangle^2$.
	\textbf{d}~Measurements of the neutral exciton $X_0$ photoluminescence saturation (crosses). The fit (red line) is used to convert pump fluency to the exciton average number $\langle N \rangle$ for each individual dot. 
	}
	\label{fig2}
\end{figure}

\subsection{Photoluminescence at high pump fluency}

The purity of single photon emission can be characterized by the second-order intensity correlation function $g_2(\tau)$, as measured by a Hanbury-Brown and Twiss (HBT) setup. In the case of pulsed excitation $g_2(\tau)$, appears as a series of peaks, whose {  spectral} width is linked to the lifetime of the different states making up the emission, as illustrated in Fig.~\ref{fig2}a. The ratio between the amplitudes of peaks at zero delay time and any of the side ones is a direct measure of the purity of the single photon emission and is denoted $g_2(0)$. Due to the limitation of our pulsed laser source, the minimum repetition rate used here is {  insufficient} to fully resolve the peaks in time, due to the extremely long lifetime of the neutral exciton, e.g. for QD16 in Fig.~\ref{fig2}a $\tau_{X_0}=354\pm19$~ns; hence $g_2(0)$ cannot be taken directly from the ratio of the peaks, but it is necessary to fit the correlation function $g_2(\tau)$, as detailed in SI.

The ability of a quantum dot to generate only one photon at a time degrades at high pump fluency, $\langle N \rangle >1$, due to the progressive contribution of biexciton and other multiexciton recombination leading to multi-photon emission. 
We here note that the number of generated electron-hole pairs $N$ follows a Poissonian distribution with {   an} average $\langle N\rangle$ proportional to the excitation power, so biexcitons can be generated even well below the saturation regime. 
In the example of Fig.~\ref{fig2}, $g_2(0)$ increases from a minimum of 0.07 for $\langle N \rangle=0.15$ to a value of 0.22 for $\langle N \rangle=1.23$. We also notice in Fig.~\ref{fig2}c a strong contribution from a very short lifetime process (cyan peaks), which we interpret as emission from spurious shell states. Such states are also typically blueshifted and exhibit classical emission properties, i.e. $g_2(0)= 1$~\cite{2Galland2012}, which further deteriorates the purity of the quantum dot emission. Although the total number of photons emitted by the shell{  --}as measured by the area of the cyan peaks{  --}is negligible compared to the core emission, their very short lifetime ($\sim$1~ns) means {  that} they contribute significantly to the measured $g_2(0)$.

\begin{figure*}
	\includegraphics[width=\linewidth]{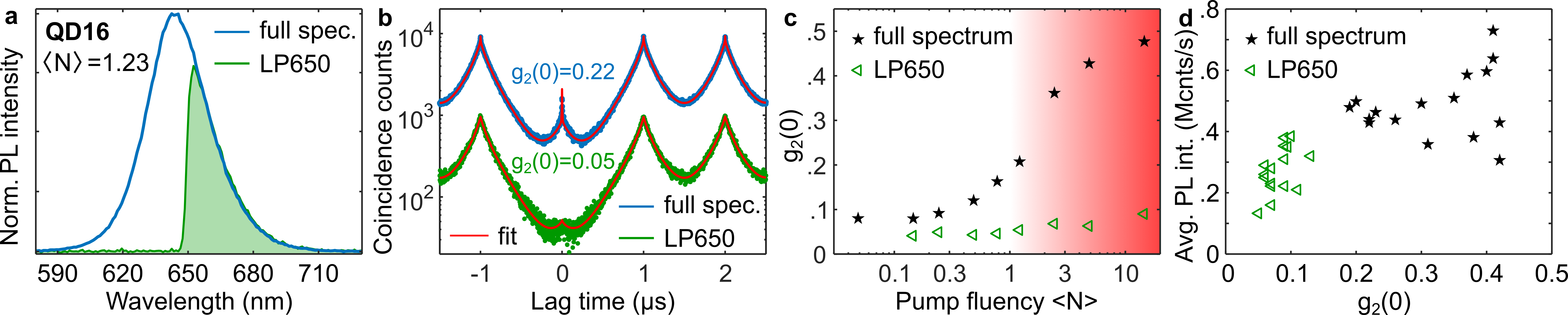}
	\caption{
	\textbf{Spectral filtering of an individual quantum dot emission with a long-pass filter at 650~nm.}
	\textbf{a}~Photoluminescence spectrum of QD16 measured at saturation (blue) and its filtered part (green-shaded area).
	\textbf{b}~Corresponding intensity correlation histograms of QD16 acquired from the entire spectrum (blue) and above 650~nm (green).
	\textbf{c}~Evolution of the single photon purity at increasing pump fluency for the entire spectrum (black stars) and the filtered part (green triangles). The saturation regime is indicated by the red-shaded area.
	\textbf{d}~Single-photon purity $g_2(0)$ and time-averaged photoluminescence brightness statistics for 16 quantum dots pumped above saturation $2 < \langle N \rangle < 5$ (black stars), and corresponding filtered emission with a long-pass filter at 650~nm (green triangles). 
	}
	\label{fig3}
\end{figure*}

The intensity correlation histograms $g_2(\tau)$ can be used to precisely estimate the relative efficiency of the biexciton emission. Indeed, it has been shown~\cite{Nair2011} that in the limit of low pump excitation, $\langle N \rangle \rightarrow 0$, the purity {  becomes} $g_2(0)\rightarrow \eta_{XX}/\eta_{X}$. {  Here,} $\eta_{XX}$ and $\eta_{X}$ are the luminescence quantum yields of biexciton and single exciton, respectively. {  The latter yield} $\eta_{X}$ can be estimated by the photoluminescence spectrally integrated counts, by accounting for the setup collection and detection efficiency and by dividing the result by the pump repetition rate. {  We note that} $g_2(0)$ is roughly constant at low power because the area of the zero-delay peak and the area of the side peaks both grow quadratically with $\langle N \rangle$ until intensity saturation is reached, as we demonstrate in Fig.~\ref{fig2}c for QD16. This is a clear proof that biexciton emission dominates the photon purity at low pump fluency~   {\cite{Park2011}}; the same analysis {  applied to} different dots shows the same quadratic behavior (see SI {Fig.~S}4).
However, as pumping approaches the exciton saturation, i.e. $\langle N \rangle \gtrapprox 1$ (red-shaded area in Fig.~\ref{fig2}c), {  contributions} of higher order multiexciton and shell emission {  are} no longer negligible, 
{hence the relative weights of peaks evolve in a complex manner.}
We note that the single photon purity at high pump fluency is not representing directly the multiexciton contribution due to significant {  contributions} of the CdS shell states with $g_2(0)=1$ (see SI {Fig.~S}5a). 

As shown in Fig.~\ref{fig2}d, the maximum brightness of the single exciton emission $X_0$ is reached around $\langle N \rangle \approx 3$, approaching a quantum yield $\eta_{X}$ of nearly 1 in this case (1~Mcounts/s for 1~MHz excitation).  Although at high pump fluency intensity blinking decreases the time-averaged brightness, due to the contribution of charged exciton states, the emission brightness remains around 50$\%$ across the ensemble of dots studied ({Fig.~S}4). The conversion between pump power and the average number of excitons generated $\langle N \rangle$ is done {\it a posteriori} for each individual dot~\cite{Park2011,Smyder2014}, by fitting the evolution of photoluminescence with pump fluency (red line in panel Fig.~\ref{fig2}d, in excellent agreement with the data). 

\begin{figure*}
	\includegraphics[width=0.9\linewidth]{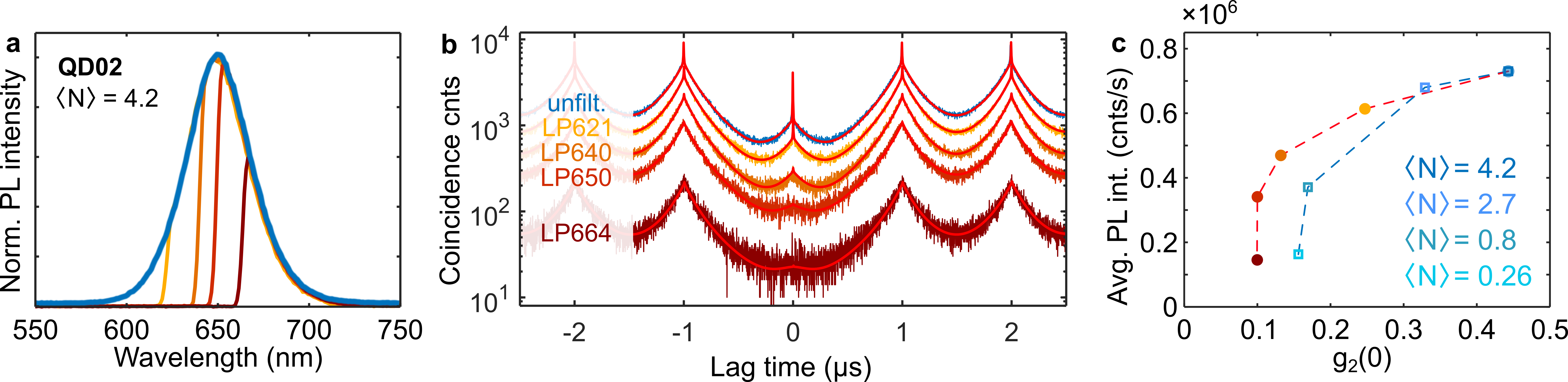}
	\caption{\textbf{Scanning photoluminescence spectrum with a tunable long-pass filter.}
	\textbf{a}~Photoluminescence spectrum of QD02 measured at high pump fluency $\langle N \rangle =4.2$ (blue) and filtered spectra at 621~nm, 640~nm, 650~nm, and 664~nm cut-off (yellow to dark red).
	\textbf{b}~Corresponding intensity correlation histograms. Solid red lines show the fit to extract $g_2(0)$.
	\textbf{c}~$g_2(0)$-intensity data pairs (circles) at $\langle N \rangle =4.2$ are extracted from single photon purification experiments shown in panels \textbf{a}-\textbf{b}. Open squares present $g_2(0)$-intensity data obtained varying pump fluency  $\langle N \rangle$ from 0.26 to 4.2. Dashed lines connecting data points serve as a guide for an eye. 
	}
	\label{tuna2}
\end{figure*}

\subsection{Single photon purification by spectral filtering}

Since the biexciton, multiexciton and CdS shell emission are all blueshifted with respect to the exciton emission, we want to study how effectively we can spectrally filter these states and isolate the single exciton emission, while at the same time preserving the quantum dot brightness. 
Fig.~\ref{fig3}a shows the emission spectrum (blue) for the same dot QD16 from Fig.~\ref{fig2}, excited at {  a} pump fluency just above saturation $\langle N \rangle =1.23$, when applying a long-pass filter with cut-off at 650~nm (green-shaded region). Removing the blue portion of the emission spectrum affects drastically the single photon purity, as demonstrated in the corresponding intensity correlation histograms in Fig.~\ref{fig3}b: $g_2(0)$ undergoes a major improvement from 0.22 (blue histogram) to 0.05 (green histogram) after filtering. Moreover, the green histogram is completely missing the very fast peak at zero delay time, showing that filtering provides excellent suppression of the short-lifetime states.

We repeat the filtering experiment in a wide range of pump fluency below and above saturation, as shown in Fig.~\ref{fig3}c. At low fluency, we can observe a 2.3 times suppression of the biexciton relative to the exciton emission, with purity improving from $g_2(0)=0.07\pm0.01$ to $g_2(0)=0.03\pm0.01$ (after background correction~\cite{Becher2001,Nakajima2012}, see SI {Fig.~S}6). This is consistent with the biexciton blueshift in this quantum dot (20~nm, see SI {Fig.~S}3) and the exciton bandwidth (33~nm). The effect of spectral filtering on the $g_2(0)$ is even larger at higher pump fluency, with a factor 6.6 observed for $\langle N \rangle=4.9$. Generally, the filtered $g_2(0)$ dependency is much flatter (green triangles in  Fig.~\ref{fig3}c), as opposed to the sharp rise observed above saturation for the unfiltered case (black stars in  Fig.~\ref{fig3}c), demonstrating an efficient suppression of higher order multiexcitons and shell state emission, in addition to the biexciton.  

We now turn to discuss the effect of filtering on photoluminescence brightness and the trade-off between purity and brightness. Evidently, spectral filtering reduces the total photoluminescence count rate.
In the case of QD16 with an emission maximum around 642~nm, the long-pass filter at 650~nm removes about 71\% of photoluminescence intensity, severely compromising the brightness of the source. As the emission wavelength changes substantially from dot to dot, some dots achieve a better balance between brightness and purity for a fixed filter cut-off at 650~nm. This is illustrated in Fig.~\ref{fig3}d, which shows a distribution of unfiltered time-averaged intensity-$g_2(0)$ data pairs for 16 quantum dots (black stars), and their corresponding filtered data (green triangles). Data are acquired for pump fluency above saturation, $2< \langle N \rangle <5$. All measurements are done at the same laser pump power, but due to the variability of absorption, $\langle N \rangle$ changes from dot to dot. 
The photoluminescence brightness is given in Mcounts/s; since the pump rate is 1~MHz, this is {  a} direct measure of the average photon rate obtainable, i.e. $0.5=50\%$ quantum yield (collection efficiency, losses and detection are all accounted for~\cite{Morozov2018}, see SI). 
On average, the brightness goes from $0.50\pm0.05$~Mcounts/s to $0.27\pm0.03$~Mcounts/s after filtering, whereas the purity improves from $g_2(0)=0.34\pm05$ to $g_2(0)=0.08\pm01$ (see more examples in SI {Fig.~S}6).
{Besides the gain in single photon purity, the spectral filtering can improve the photo-stability of emission as spurious states are removed from the signal. However, it may also introduce additional fluctuations of the bright excitonic state due to an inappropriate choice of filtering wavelength in combination with spectral diffusion at room temperature (see SI    {Fig.~S}2 and    {Fig.~S}8b).}
This analysis clearly shows the need to search for an optimal cut-off wavelength for spectral filtering, which can balance single photon purification with tolerable losses    {and photo-stability}.

\subsection{Purity-brightness trade-off}

We employed a tunable long-pass filter to study the optimal cut-off wavelength for a purity-brightness trade-off. 
Fig.~\ref{tuna2}a presents a photoluminescence spectrum of an individual quantum dot labeled QD02 (blue line) excited at high pump fluency $\langle N \rangle =4.2$, and the filtered portions of its spectrum for cut-off wavelengths in the 621--664~nm range of the long-pass filter (yellow to dark red). 
Fig.~\ref{tuna2}b shows the corresponding intensity correlation histograms.  
The unfiltered histogram (blue) has a $g_2(0)= 0.44$ and exhibits a fast decay peak at zero correlation time, indicating substantial contribution of classical emission from the CdSe shell, as well as higher order multiexcitons, which significantly compromise its single photon purity.
Filtering at 621~nm in Fig.~\ref{tuna2}b (yellow curve) starts to suppress the fast sub-nanosecond component, without, however, removing the slower lifetime peak at zero delay, which we attribute mainly to the biexciton contribution.
A 640~nm cut-off completely removes the fast component and significantly reduce the central peak (orange curve). 
Shifting the filtering wavelength above 650~nm further decreases the biexciton contribution (red), whereas a cut-off of 664~nm (dark red) mainly suppresses the number of counts without significantly reducing the central peak relative height, demonstrating {  that} the exciton emission is heavily filtered at this stage. 
We summarize these experiments in Fig.~\ref{tuna2}c, where we plot photoluminescence brightness versus $g_2(0)$ (circles); the color-coding is the same as in {  panels} a and b {  while} points are connected by a dashed line as a guide for {  the eyes} (see SI {Fig.~S}8a for more single dot examples). It is clear that the largest purity improvement happens for a cut-off of 640~nm, with $g_2(0)$ going from 0.44 to 0.13, while causing only a $35\%$ reduction of photoluminescence brightness. We compare these single photon purification results with $g_2(0)$-intensity values obtained by varying pump fluency (open squares in Fig.~\ref{tuna2}c).  At large pump fluency $\langle N \rangle >1$, the effect of spectral filtering is comparable to just reducing the pump power; however, for pump fluency close to saturation, $\langle N \rangle \approx 1$, the spectral filtering clearly outperforms the technique based on pump power. For instance, one can obtain the same brightness as  $\langle N \rangle =0.8$ when pumping at $\langle N \rangle =4.2$ with a 650~nm long-pass filter, while significantly improving the purity of the emission. The fluency threshold at which purification based on spectral filtering overcomes the simple power reduction is different {  from} dot to dot, but in all of {  the} quantum dots under study we found an improvement. Choosing the right cut-off wavelength for each dot is essential to guarantee the best purity-brightness trade-off.

Long exciton lifetime reported for this class of quantum dots could be considered a limitation to the maximum achievable brightness. However, recent studies on similar dots have highlighted that lifetimes can be reduced by working with highly charged excitons, while maintaining low blinking and large quantum yields~\cite{Morozov2020}. Moreover, coupling with plasmonic and dielectric nanoantennas could also be used to reduce the radiative lifetime and improve brightness, without affecting the spectral {  positions} of bi- and multiexcitons.

\section{Conclusion}
In conclusion, we reported a large biexciton spectral shift of $24\pm5$~nm, at the single dot level, for pure-phase wurtzite CdSe/CdS giant shell quantum dots with long lifetime and high quantum yield. We demonstrated that spectral filtering can be effectively used in these quantum dots to improve the emission purity at room temperature.
We showed that spectral filtering using a long-pass filter can improve single photon purity both at low and high excitation fluency. At low powers, $g_2(0)$ improves by as much as factor of 2.3 (to $g_2(0)=0.03\pm0.01$), showing a promising suppression of biexciton emission, which can be controlled by tuning the cut-off of the filter. When pumping just above saturation, to ensure high brightness and deterministic emission, the spectral purification is even more effective, due to the effective suppression of multiexciton and shell emission. 
The biexciton spectral separation in giant shell quantum dots can be further increased through the engineering of strain at {  the} core/shell interface~\cite{Polovitsyn2018,Bae2013}, which, in combination with the filtering technique explored here, would result in bright and pure single photon sources at room temperature, opening new routes for applications in quantum cryptography and communications.

\bibliography{bibliography}
\bibliographystyle{unsrt}

\section{Supporting Information}
Supporting Information includes details on synthesis of giant shell quantum dots;
 photoluminescence measurements;
 fitting intensity correlation histograms; emission statistics of individual quantum dots; photoluminescence spectra of QD16 and QD02 at high pump fluency; evolution of the central and side peaks in intensity correlation histogram with pump fluency; emission of QD16 at increasing pump fluency; background emission at increasing excitation power; other examples of scanning a quantum dot spectrum with a tunable long-pass filter, and of filtering with a long pass filter at 650~nm. 
The data that support the findings of this study are available from the corresponding author upon reasonable request.

\section{Acknowledgments}

S.~M. acknowledges funding from the Marie Sk\l{}odowska-Curie Action (grant No.~101032967).
N.~A.~M. is a VILLUM Investigator supported by VILLUM FONDEN (grant No.~16498).
I.~M., A.~M. and A.~D.~G. acknowledge funding from the European Research Council (ERC) under the European Union’s Horizon 2020 research and innovation program (grant No. 714876 PHOCONA), and the Research Foundation -- Flanders (grant No. G037221N HITEC).
R.S. and S.V. acknowledge funding from the Engineering and Physical Sciences Research Council (EP/V048880 and EP/P033431).

\end{document}